\documentclass[aps,showpacs,pra,twocolumn,superscriptaddress]{revtex4}%
\usepackage{multirow}
\usepackage{color}
\usepackage{amsfonts}
\usepackage{amsmath}
\usepackage{mathrsfs}
\usepackage{amssymb}
\usepackage{wasysym}
\usepackage{graphicx}
\begin{document}
\title{Generalized Effective Potential Landau Theory for Bosonic Superlattices}
\author{Tao Wang}
\email{tauwaang@gmail.com} \affiliation{Department of Physics,
Harbin Institute of Technology, Harbin 150001, China}
\affiliation{Physics Department and Research Center OPTIMAS,
University of Kaiserslautern, 67663 Kaiserslautern, Germany}
\author{Xue-Feng Zhang}
\affiliation{Physics Department and Research Center OPTIMAS,
University of Kaiserslautern, 67663 Kaiserslautern, Germany}
\author{Sebastian Eggert}
\affiliation{Physics Department and Research Center OPTIMAS,
University of Kaiserslautern, 67663 Kaiserslautern, Germany}
\author{Axel Pelster}
\email{axel.pelster@physik.uni-kl.de} \affiliation{Physics
Department and Research Center OPTIMAS, University of
Kaiserslautern, 67663 Kaiserslautern, Germany}
\affiliation{Hanse-Wissenschaftskolleg, Lehmkuhlenbusch 4, 27733
Delmenhorst, Germany}
\begin{abstract}
We study the properties of the Bose-Hubbard model for
square and cubic superlattices. To this end we
generalize a  recently established effective potential Landau
theory for a single component to the case of multi components and
find not only the characteristic incompressible solid phases with
fractional filling, but also obtain the underlying quantum phase
diagram in the whole parameter region at zero temperature.
Comparing our analytic results with corresponding ones from
quantum Monte Carlo simulations demonstrates the high accuracy of
the generalized effective potential Landau theory (GEPLT).
Finally, we comment on the advantages and disadvantages of the
GEPLT in view of a direct comparison with a corresponding
decoupled mean-field theory.
\end{abstract}
\maketitle

\section{Introduction}
Systems of ultracold bosonic gases in optical lattices have
recently become a major field in physics
research \cite{Lewenstein:AdP07,Bloch:RMP08,Sanpera}. After their
theoretical suggestion \cite{Fisher:PRB89,Jaksch:PRL98} and first
experimental realization
using counter-propagating laser beams \cite{Greiner:Nat02} it soon became clear
that they establish a versatile bridge between the field of
ultracold quantum matter and correlated
condensed matter systems \cite{Bloch:NaP12}.

One of the most famous examples is the Bose-Hubbard model
\cite{Fisher:PRB89,Jaksch:PRL98}  which undergoes a quantum phase
transition from a Mott insulator to a superfluid phase due to the
competition between the atom-atom on-site interaction and the
hopping amplitude.   This transition can be demonstrated
experimentally by time-of-flight absorption
pictures \cite{Greiner:Nat02}, or measuring the collective
excitation spectra via Bragg
spectroscopy \cite{sengstock1,sengstock2}. Recent research efforts
have targeted more complex systems, which include long-range
interactions (e.g. from dipolar
bosons \cite{Fleischhauer1,bloch:na12}), mixtures of several
components \cite{Hofstetter1,sengstock3} and more interesting
lattice geometries, such as frustrated or superlattice
structures \cite{experiment1,experiment2,experiment4,experiment3,experiment5}.
Accordingly, the corresponding phase diagrams become richer and
more complex, including the possibility of phases with periodic density modulations or supersolidity.
A crystalline density wave
phase, for instance, generally occurs at fractional filling and it has been
proposed that the corresponding commensurate density modulation could be
detected by measuring correlations with time-of-flight and
noise-correlation techniques \cite{hou}. Furthermore, recent
experimental progress in achieving single-site addressability in
optical lattice structures \cite{ss1,ss2,ss3,ss4,ss5,ss6,ss7}
nourishes the prospect to directly observe density wave modulations in
the near future.

\begin{figure}[t]
\includegraphics[width=0.5\textwidth]{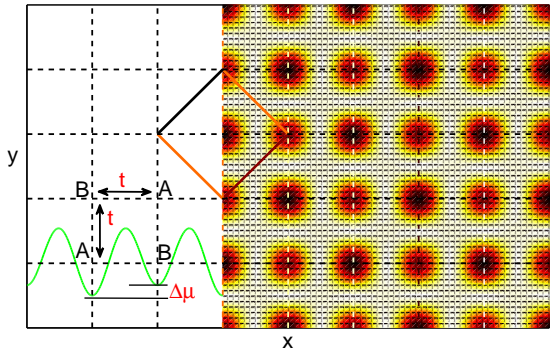}
\caption{(Color online) Schematic illustration for a
square superlattice in two dimensions. The solid line
square represents one type of the unit cell. The solid curve
(green)  shows the optical potential in the $x$ direction. Lattice
sites $A$ are deeper by $\Delta \mu$ than lattice sites $B$.}
\label{figure1}
\end{figure}

Such density modulations may emerge from interactions via spontaneous symmetry-breaking, but a
a simpler way to create them is with
a superlattice generated from commensurate
lasers, see for instance Refs. \cite{experiment1,experiment2,experiment4,experiment3,experiment5} for further experimental details.
Thus, then the potential depth is slightly different on one
sublattice, while the interaction strength and hopping amplitude
will remain almost uniform. The corresponding Bose-Hubbard model
Hamiltonian on a square or cubic lattice is given by \cite{vezzani}
\begin{eqnarray}
& & \hspace{-3mm}\hat{H}_{\rm SL} =-t\sum_{\langle j\in A, j'\in
B\rangle}\left(\hat{a}_j^{\dag}\hat{a}_{j'}^{\phantom{\dag}}+\hat{a}_j^{\phantom{\dag}}\hat{a}_{j'}^{\dag}\right) \label{bh}\\
&&
\nonumber +\frac{U}{2} \sum_{j\in
A,B}\hat{n}_j\left(\hat{n}_j-1\right)-\left(\mu+\Delta\mu\right)
\sum_{j\in A}\hat{n}_j-\mu \sum_{j\in B}\hat{n}_j  \, ,
\end{eqnarray}
where $\Delta \mu$ stands for a small additional chemical
potential on sublattice $A$ compared to sublattice B, as
illustrated in Fig.~\ref{figure1}. As we will show this model
exhibits an interesting competition between Mott and density wave
phases.

From a theoretical point of view the study of interacting bosons
and quantum phase transitions is far from trivial \cite{Sachdev}.
The possible phases in different kinds of optical superlattices
have so far been analyzed by numerical approaches
\cite{scalettar,schollwoeck,das2}, decoupled mean field
theory \cite{vezzani,hou,shuchen,das1}, multisite mean-field theory \cite{msmf1,msmf2,msmf3,msmf4} and cell
strong-coupling expansion method \cite{csce1,csce2}. The latter method yields excellent
results for $1$d systems when compared to the powerful
numerical method of Ref.~\cite{msmf3}.
However, it is known that mean-field theory can have
significant deviations from unbiased high-precision numerical
results \cite{zhang} and the strong-coupling expansion is not that
accurate when applied to higher dimensional systems. The purpose
of this paper is, therefore, to present a reliable quantitative method to
determine non-trivial phases of high-dimensional
multi-component boson systems. To this end we profit from recent advances to use a
systematic Landau theory with an effective potential that can be
estimated quantitatively from the microscopic model, e.g.~by
diagrammatic
methods \cite{pelster1,pelster2,pelster3,pelster4,jiangying,pelster5}.
Whereas the first hopping order of the effective potential Landau
theory leads to similar results as mean-field
theory \cite{Fisher:PRB89}, higher hopping orders have recently
been evaluated via the process-chain
approach \cite{Eckardt,Holthaus5,Holthaus6,Teichmann,Holthaus7},
which determines the location of the quantum phase transition for
the single component Hubbard model for cubic as well as triangular
and hexagonal optical lattices to a similar precision as demanding
quantum Monte Carlo simulations \cite{MC1,MC2}. Thus, it becomes
even possible to calculate the critical exponents of the
corresponding quantum phase
transition \cite{exponents1,exponents2}.  We now present a
generalized effective potential Landau theory (GEPLT), which
extends those concepts to multi-component systems and to phases
with non-trivial crystalline order parameters.  In particular, for
the model in Eq.~(\ref{bh}) the GEPLT approach gives excellent
quantitative estimates for the location of the phase boundaries
compared to unbiased quantum Monte Carlo simulations.

At first, we briefly review the effective potential Landau theory
for the single component Bose-Hubbard model in Sec.~\ref{LT}.
Then, we extend this
method step by step from one component to the general superlattice case in
Section \ref{MC}. After that, we apply this GEPLT method to
the simple superlattice model in Eq.~(\ref{bh})
and determine
the resulting quantum phase diagram at zero temperature in the whole parameter region
in Section \ref{OS}. Both the advantages and disadvantages of
GEPLT are revealed by comparing it with a decoupled
mean-field theory in Section \ref{AD}. Finally, Section \ref{CO}
provides the conclusions and sketches related problems in an outlook.
\section{Effective Potential Landau Theory}\label{LT}

Let us first consider the
Bose Hubbard model in Eq.~(\ref{bh})
for the
well-studied case of $\Delta\mu=0$ \cite{Fisher:PRB89,Jaksch:PRL98}.
The second-order quantum phase transition between
the Mott insulator, which occurs for $t/U \ll 1$, and the
superfluid, which is realized for $t/U \gg 1$, is intimately
connected with a spontaneous breaking of the underlying
U(1)-symmetry of the Bose-Hubbard model (\ref{bh}). To describe
this theoretically, we transfer the usual field-theoretic approach
for thermal phase transitions \cite{Zinn-Justin,Kleinert} to quantum phase
transitions and couple the creation and
annihilation operators to external source fields with uniform
strength $J$ and $J^{*}$ within a Landau theory
\cite{pelster2,pelster3}
\begin{eqnarray}
\hat{H}_{\rm BH} (J,J^*)=\hat{H}_{\rm BH} + \sum_j \left( J \hat{a}_{j}^{\dag} + J^* \hat{a}_{j} \right) \, .
\label{BH2}
\end{eqnarray}
The transition from the Mott insulator to the superfluid phase is
described by the emergence of a non-vanishing order parameter which is defined due to homogeneity according to
$\psi = \langle \hat{a}_i\rangle$, $\psi^* =  \langle
\hat{a}_i^{\dag}\rangle$. The free energy corresponding to (\ref{BH2})
\begin{eqnarray}
F(J,J^{*}) = -\frac{1}{\beta} \, \ln \left[ \mbox{Tr}\, e^{- \beta \hat{H}_{\rm BH} (J,J^*)}
\right]
\label{BH3}
\end{eqnarray}
allows to determine this order parameter via
\begin{eqnarray}
\psi = \frac{1}{N_{s}}\frac{\partial
F(J,J^{*})}{\partial J^{*}}\,, \quad
\psi^* =  \frac{1}{N_{s}}\frac{\partial
F(J,J^{*})}{\partial J}\,,
\label{psi}
\end{eqnarray}
where $N_{s}$ denotes the number of lattice sites.  Equation (\ref{psi}) motivates
that it is possible to formally perform a Legendre transformation from the free
energy $F(J, J^{*})$ in
order to arrive at an effective potential $\Gamma(\psi, \psi^*)$
that is useful in a quantitative Landau theory
\begin{eqnarray}
\label{L}
\Gamma(\psi,\psi^{*})=F/N_{s}-\psi^{*}J-\psi J^{*}\,.
\end{eqnarray}
Due to Legendre identities the external sources can be reobtained from derivatives of the
effective potential
\begin{eqnarray}
\label{J}
\frac{\partial \Gamma}{\partial \psi^{*}}=-J, \quad \frac{\partial
\Gamma}{\partial \psi}=-J^{*}\,.
\end{eqnarray}
The original Bose-Hubbard Hamiltonian (\ref{bh}) is restored from
(\ref{BH2}) for vanishing currents, i.e. by setting $J=J^*=0$. In this limit we conclude from
(\ref{L}) that the effective potential reduces to the free energy.
Furthermore, Eq.~(\ref{J}) then implies that the order parameter $\psi,
\psi^*$ of the system follows from extremizing the effective
potential. A trivial extremum $\psi=0$ corresponds to the
Mott-insulator phase, whereas a non-vanishing extremum $\psi \neq
0$ occurs in the superfluid phase.

The free energy (\ref{BH2}) reduces at zero temperature to the ground-state energy, which can be calculated
in a power series of
both the hopping parameter $t$ and the source terms $J, J^{*}$ by using
the Rayleigh-Schr\"{o}dinger perturbation
theory \cite{pelster1,pelster2,pelster3,pelster4,jiangying,pelster5,Eckardt,Holthaus5,Holthaus6,Teichmann,Holthaus7}.
Due to the underlying U(1)-symmetry of the Bose-Hubbard
Hamiltonian (\ref{bh}) the expansion is only a power series in terms of $|J|^2$
\begin{eqnarray}
\label{free}
F(J,J^{*},t)=N_{s}\left(F_{0}(t)+\sum_{p=1}^{\infty}
c_{2p}(t)\left|J \right|^{2p}\right)\, ,
\end{eqnarray}
where the respective
expansion coefficients are accessible via a hopping expansion
\begin{eqnarray}
c_{2p}(t)=\sum_{n=0}^{\infty} (-t)^{2n} \alpha_{2p}^{(n)}\,.
\end{eqnarray}
From Eqs.~(\ref{psi}), (\ref{J}), and (\ref{free}) we then obtain the effective potential of the Bose-Hubbard
Hamiltonian (\ref{bh}) in the following perturbative form
\begin{eqnarray}
\Gamma(\psi,\psi^{*},t)=F_{0}(t)-\frac{1}{c_{2}(t)}|\psi|^2+\frac{c_{4}(t)}{c_{2}(t)^{4}}|
\psi|^4+\cdots\,.
\label{gamma}
\end{eqnarray}
According to the Landau theory for second-order phase transitions, the
critical line between the Mott insulator and the
superfluid phase follows from finding the zero of the second-order
coefficient in (\ref{gamma}). In order to solve the resulting
equation $1 / c_{2}(t_c) =0$, we expand it in a power series of
the hopping parameter $t$
\begin{eqnarray}
\frac{1}{c_{2}(t)}&=&\frac{1}{\alpha_{2}^{(0)}}
\left(1+\frac{\alpha_{2}^{(1)}}{\alpha_{2}^{(0)}}t+\left[\left(\frac{\alpha_{2}^{(1)}}{\alpha_{2}^{(0)}}\right)^{2}
-\frac{\alpha_{2}^{(2)}}{\alpha_{2}^{(0)}}\right]t^{2} \right. \nonumber \\
& & +\cdots\Big)\,.
\label{SCHE}
\end{eqnarray}
Thus this gives us an algebraic equation for $t_{c}$, whose degree
depends on the respective hopping order which is taken into
account. The number of the roots is the same as the order of $t$,
but only the smallest real positive root is identified as an
appropriate approximation for the location of the quantum phase
transition.

As mentioned in the introduction, the effective potential
Landau theory was quite successful in calculating the quantum
phase boundary for the single component
system \cite{pelster1,pelster2,pelster3,pelster4,jiangying,pelster5,Eckardt,Holthaus5,Holthaus6,Teichmann,Holthaus7}.
However, it cannot
be used to treat a superlattice system, since more than one
order parameter appears. Therefore, we will work out in the next
section a corresponding extension to multi components which
overcomes this problem.
\section{Generalized Effective Potential Landau Theory}\label{MC}
In order to describe a superlattice or a multi-component system, we have to
introduce several sites or degrees of freedom at each lattice point.
In other words, we introduce a larger unit cell at each lattice point, labeled by $j$,
together with a basis of size $m$, labeled by $l=1,..,m$.  The generalized
Bose-Hubbard Hamiltonian with $m$ bosonic species in each unit
cell is therefore given by
\begin{eqnarray}
\hat{H}_{\rm BH} &=&-
 \sum_{j,j'}
\sum_{l,l'=1}^{m}
\left[ t_{j(l),j'(l')} \hat{a}_{j(l)}^{\dagger}\hat{a}_{j'(l')}^{\phantom{\dag}} + \mbox{h.c.} \right] \\
&&+\sum_{j}\sum_{l=1}^{m}\left[\frac{U_{(l)}}{2}\hat{n}_{j(l)}\left(\hat{n}_{j(l)}-1\right)
-
\mu_{(l)}\hat{n}_{j{(l)}}\right] \, , \nonumber
\end{eqnarray}
where $\hat{a}_{j(l)}$ denotes the boson annihilation operator at
lattice point $j$ with basis index $(l)$.   Hopping
$t_{j(l),j'(l')}$ can occur between any basis and lattice
position, while the repulsion $U_{(l)}$ acts for now only between
bosons of the same lattice point and basis index.  The chemical
potential $\mu_{(l)}$ depends on the basis index which is
analogous to sublattices $A$ and $B$ in Eq.~(\ref{bh}).

We now model the symmetry-breaking by introducing the source
vectors $\vec{\mathbb{J}}=(J_{1},\ldots,J_{m})^{\rm T}, \vec{
\mathbb{J}}^{\dag}=(J_{1}^*,\ldots,J_{m}^*)$ according to
\begin{eqnarray}
\hat{H}_{\rm BH} (\vec{\mathbb{J}}, \vec{\mathbb{J}}^{\dag}) =
\hat{H}_{\rm BH} +\sum_{j} \sum_{l=1}^m \left( J_{l}^{\phantom{*}}
\hat{a}_{j{(l)}}^{\dagger}+J_{l}^{*}
\hat{a}_{j{(l)}}^{\phantom{\dag}}\right) \, .
\end{eqnarray}
By generalizing the procedure from a single component to multi
components, we use perturbation theory in order to determine the
free energy at zero temperature in a power series of both the
hopping parameters $t_{j{(l)},j'{(l')}}$  and the source vectors
$\vec{\mathbb{J}}, \vec{\mathbb{J}}^{\dag}$.  In principle, we
need an expansion in terms of all relevant hopping parameters
$t_{j{(l)},j'{(l')}}$, but to illustrate the process we consider
here the case that only one hopping element $t$ dominates
(e.g.~between nearest neighbors) and all others are neglected
\begin{eqnarray}
F(\vec{\mathbb{J}},\vec{\mathbb{J}}^{\dag},t)= N_{s}\Big[
F_{0}(t)+\vec{\mathbb{J}}^{\dag} C_2(t) \vec{\mathbb{J}} +\cdots
\Big] \,. \label{second}
\end{eqnarray}
The matrix elements $c_{2ll'}^{\phantom{(n)}}(t)$ of $C_2(t)$ are then
given by a hopping expansion of the form
\begin{eqnarray}
\label{CST}
c_{2ll'}^{\phantom{(n)}}(t)=\sum_{n=0}^{\infty}\left(-t\right)^{n}\alpha_{2ll'}^{(n)}\,.
\end{eqnarray}
The order parameter vectors give different values for each basis
index, but are independent of lattice points
$\vec{\Psi}=(\psi_1,\ldots,\psi_m)^{\rm T},
\vec{\Psi}^{\dag}=(\psi_1^*,\ldots,\psi_m^*)$ according to
\begin{eqnarray}
\vec{\Psi}&=&
\left( \langle \hat{a}_1\rangle,\ldots,\langle \hat{a}_m\rangle \right)^{\rm T} \, , \nonumber \\
\label{psiv}
\vec{\Psi}^{\dag}&=& \left( \langle
\hat{a}_1^{\dag}\rangle,\ldots,\langle \hat{a}_m^{\dag}\rangle
\right) \,,
\end{eqnarray}
and we observe
\begin{eqnarray}
\psi_i=\frac{1}{N_{s}} \frac{\partial F}{\partial
J^*_i} \,, \quad
\psi_i^*=\frac{1}{N_{s}} \frac{\partial F}{\partial
J_i } \,. \label{psiva}
\end{eqnarray}
Again this motivates to perform the Legendre transformation of the
free energy.  The generalized effective potential then depends on
the order parameter vectors $\vec{\Psi}, \vec{\Psi}^{\dag}$:
\begin{eqnarray}
\label{Gamma}
\Gamma(\vec{\Psi},\vec{\Psi}^{\dag},t)=F/N_{s}-\vec{\mathbb{J}}^{\dag}\vec{\Psi}-\vec{\Psi}^{\dag}\vec{\mathbb{J}}\,.
\end{eqnarray}
Legendre identities allow to write the external sources as derivatives of the
effective potential
\begin{eqnarray}
\label{Jv}
\frac{\partial \Gamma}{\partial \psi_i}=- J_i^*\,,\quad
\frac{\partial \Gamma}{\partial \psi_i^*}=-J_i \,,
\end{eqnarray}
so the order parameter vector is determined by extremizing
$\Gamma$ in the physical limit that the external source vectors
$\vec{\mathbb{J}}, \vec{\mathbb{J}}^{\dag}$ vanish.

Due to Eqs.~(\ref{second}), (\ref{psiva}), and (\ref{Gamma}) the effective potential of the system is of the form
\begin{eqnarray}
\label{ga}
\Gamma(\vec{\Psi},\vec{\Psi}^{\dag},t)=F_{0}(t)+\vec{\Psi}^{\dag}
A_2 (t)\vec{\Psi} + \cdots\,.
\end{eqnarray}
The resulting relation between the matrices $A_2$ and $C_2$ can be deduced in
following way. By inserting (\ref{ga}) into (\ref{Jv}), we get
\begin{eqnarray}
A_{2.ij}
=\frac{\partial^2\Gamma}{\partial \psi_i^*  \partial \psi_j}=-\frac{\partial J_i}{\partial \psi_j}\,.
\end{eqnarray}
Combining this with (\ref{second}) and (\ref{psiva}), we read off
\begin{equation}
-\delta_{ij}= A_{2,ik} \frac{\partial \psi_k}{\partial J_j}= \frac{1}{N_{s}} A_{2,ik}
\frac{\partial^2 F}{\partial J_k^* \partial J_j}=\left( A_2C_2\right)_{ij} \, . \label{inverse}
\end{equation}
Thus, the matrix $A_2$ turns out to be the inverse of $-C_2$. As
all matrix elements of $C_2$ are given by a hopping expansion of the form (\ref{CST}),
we get a corresponding hopping expansion for each element of $A_2$.
In matrix form the first terms of this hopping expansion read
\begin{eqnarray}
&&\hspace*{-4mm}C_{2}^{-1}=\left( \alpha_{2}^{(0)}\right)^{-1}
\left\{1+\alpha_{2}^{(1)}\left(\alpha_{2}^{(0)}\right)^{-1}t +\left[\alpha_{2}^{(1)} \left(\alpha_{2}^{(0)}\right)^{-1}
\right. \right. \nonumber \\
&&\hspace*{-4mm}\left. \left.
\alpha_{2}^{(1)}\left(\alpha_{2}^{(0)}\right)^{-1}
+\alpha_{2}^{(2)}\left(\alpha_{2}^{(0)}\right)^{-1}\right]t^{2}+...\right\}
\, ,
\end{eqnarray}
which reduces for a single component to (\ref{SCHE}). The critical
line, where the order parameter vector $\vec{\Psi}$ changes from
zero to non-zero, follows then from extremizing the effective
potential (\ref{ga}). In case that all components of the order
parameter vector $\vec{\Psi}$ are non-zero, we obtain
\begin{eqnarray}
\label{DET}
\mbox{Det}\, A_2 =0\,.
\end{eqnarray}
But it could also happen that only a subset of components of the
order parameter vector $\vec{\Psi}$ is non-vanishing, which yields
the condition that the determinant of the corresponding submatrix
of $A_2$ vanishes. The physically realized quantum phase boundary
corresponds then to the smallest value of the hopping parameter
$t$, which follows from all these conditions. In the next section
we will study along these lines the most simple case of a
superlattice system which is provided by bosons on a
square or cubic  superlattice given in Eq.~(\ref{bh}).
\section{Square and Cubic Superlattice}\label{OS}
Similar to the continuous translational symmetry breaking
artificially introduced by the optical lattice to mimic a real
crystal, the optical superlattice can break the discrete
translational symmetry to study the multi-components system.
Beside that, it also can be used as a platform for
disorder \cite{disorder} and topological order \cite{TI1,TI2}
problems. Here, we apply the GEPLT to the simple square
and cubic case.
\subsection{Application of Effective Potential Theory}
Following the GEPLT from the previous section for the model in
Eq.~(\ref{bh}) we need to use for the two sublattices two
independent source terms $\vec{\mathbb{J}}=(J_A,J_B)^{\rm T}$,
yielding
\begin{eqnarray}
\hat{H}_{\rm SL} (\vec{\mathbb{J}},
\vec{\mathbb{J}}^{\dag})&=&\hat{H}_{\rm SL}
+\sum_{j \in A}\left(J_A\hat{a}_j^{\dag}+J_A^*\hat{a}_j^{\phantom{\dag}}\right) \nonumber \\
&& +\sum_{j \in B}\left(J_B\hat{a}_j^{\dag}+J_B^*\hat{a}_j^{\phantom{\dag}}\right)\,.
\label{bh2}
\end{eqnarray}
The free energy of the system can then be written as (\ref{second}),
where $C_2(t)$ represents a 2x2-matrix with the following hopping expansion
\begin{eqnarray}
\left(%
\begin{array}{cc}
  c_{2AA} & c_{2AB} \\
  c_{2BA} & c_{2BB} \\
\end{array}%
\right)=\sum_{n=0}^{\infty}\left(-t\right)^n\left(%
\begin{array}{cc}
  \alpha_{2AA}^{(n)} & \alpha_{2AB}^{(n)} \\[1mm]
  \alpha_{2BA}^{(n)} & \alpha_{2BB}^{(n)} \\
\end{array}
\right)\, ,
\label{relation}
\end{eqnarray}
where the symmetry $c_{2AB}=c_{2BA}$ holds. Then, after the
Legendre transformation (\ref{Gamma}), we obtain the effective
potential (\ref{ga}), where we have $\vec{\Psi}=(\langle
\hat{a}_A\rangle,\langle \hat{a}_B\rangle)^{\rm T}$ and $A_2$ is
the inverse of $-C_2$ according to Eq.~(\ref{inverse}). When the
second-order quantum phase transition occurs, the vanishing order
parameter vector $\vec{\Psi}=(0,0)^{\rm T}$ changes from stable to
unstable. It turns out that the smallest critical hopping
parameter results from the condition (\ref{DET}) that the
determinant of $A_2$ vanishes. With this we obtain in second order
of $t$ the following equation for the location of the quantum
phase boundary
\begin{eqnarray}
(\beta^{(0)})^2-\beta^{(0)}\beta^{(1)}t-\frac{\beta^{(2)}t^{2}}{2}=0 \, ,
\end{eqnarray}
where the abbreviations $\beta^{(0)}=\sqrt{\alpha_{2AA}^{(0)}\alpha_{2BB}^{(0)}}$,
$\beta^{(1)}= \alpha_{2AB}^{(1)}$, and
$\beta^{(2)}=\alpha_{2AA}^{(2)}\alpha_{2BB}^{(0)}+\alpha_{2AA}^{(0)}\alpha_{2BB}^{(2)}
-2(\alpha_{2AB}^{(1)})^{2}$ have been introduced. Taking into account the smallest root then yields
\begin{eqnarray}
\label{tc}
t_{c}=\frac{\beta^{(0)}\left[-\beta^{(1)}+\sqrt{(\beta^{(1)})^2+2\beta^{(2)}\,}\right]}{\beta^{(2)}}\,.
\end{eqnarray}
Thus, the problem of finding the quantum phase boundary has been
reduced to the calculation of the perturbative coefficients
$\alpha_{2ll'}^{(n)}$ in the respective hopping order. According to Appendix \ref{AA}
this perturbative calculation can be systematically performed by using a
suitable diagrammatic representation.  We use the unperturbed energies
\begin{eqnarray}
E^{(0)}\left(n_{A},n_{B}\right)&=&
\frac{U}{2}n_{A}\left(n_{A}-1\right)+\frac{U}{2}n_{B}\left(n_{B}-1\right)\nonumber \\
&&-\left(\mu+\Delta \mu \right)n_{A} -\mu n_{B}\, ,
\label{E}
\end{eqnarray}
to define the energy differences between different particle number sectors
\begin{eqnarray}
\lambda_{A}^{\pm
(p)}=&\left[ E^{(0)}\left(n_{A},n_{B}\right)-E^{(0)}\left(n_{A}\pm p,n_{B}\right) \right]/N_s,\nonumber \\
\lambda_{B}^{\pm (p)}=&\left[E^{(0)}\left(n_{A},n_{B}\right)-E^{(0)}\left(n_{A} ,n_{B}\pm
p\right)\right]/N_s.
\label{P}
\end{eqnarray}
For $p=\pm 1$ the short notation
$\lambda_{A}^{\pm }=\lambda_{A}^{\pm (1)}$, $\lambda_{B}^{\pm
}=\lambda_{B}^{\pm (1)}$ is used.
In zeroth and first hopping order we obtain the following results
for the respective coefficients $\alpha^{(n)}_{2ll'}$
\begin{eqnarray}
\alpha_{2AA(BB)}^{(0)}=\frac{n_{A(B)}+1}{\lambda_{A(B)}^{+}}+\frac{n_{A(B)}}{\lambda_{A(B)}^{-}}
\label{0}\\
\alpha_{2AB}^{(1)}=\alpha_{2BA}^{(1)}=z\alpha_{2AA}^{(0)}\alpha_{2BB}^{(0)}\, ,
\label{1}
\end{eqnarray}
whereas in second hopping order we get
\begin{eqnarray}
\nonumber
&&\alpha_{2AA}^{(2)}=z(z-1)(\alpha_{2AA}^{(0)})^2\alpha_{2BB}^{(0)}
+z\left[\frac{n_{A}^{2}n_{B}}{{\left(\lambda_{A}^{-}\right)^{2}\lambda_{B}^{-}}}\right.
\nonumber\\
&&+\frac{n_{A}n_{B}\left(1+n_{A}\right)}{\lambda_{A}^{-}\left(\lambda_{A}^{+}+\lambda_{B}^{-}\right)}
\left(\frac{2}{\lambda_{B}^{-}}-\frac{1}{\lambda_{A}^{+}+\lambda_{B}^{-}}-\frac{1}{\lambda_{A}^{-}}
\right)
\nonumber \\
&&+\frac{n_{A}\left(1+n_{A}\right)\left(1+n_{B}\right)}{\lambda_{A}^{+}\left(\lambda_{A}^{-}+\lambda_{B}^{+}\right)}\left(
\frac{2}{\lambda_{B}^{+}}
-\frac{1}{\lambda_{A}^{-}+\lambda_{B}^{+}}
-\frac{1}{\lambda_{A}^{+}}\right)
\nonumber\\
&&+\frac{n_B(n_A+1)^2}{\lambda_{A}^{+}+\lambda_{B}^{-}}\left(
\frac{\lambda_{A}^{+}-\lambda_{B}^{-}}{\lambda_{A}^{+}\lambda_{B}^{-}\left(\lambda_{A}^{+}+\lambda_{B}^{-}\right)}
-\frac{1}{(\lambda_{A}^{+})^2}\right)
\nonumber \\
&&+\frac{n_{A}^{2}\left(1+n_{B}\right)}{\lambda_{A}^{-}+\lambda_{B}^{+}}\left(
\frac{\lambda_{A}^{-}-\lambda_{B}^{+}}{\lambda_{A}^{-}\lambda_{B}^{+}\left(\lambda_{A}^{-}+\lambda_{B}^{+}\right)}
-\frac{1}{\left(\lambda_{A}^{-}\right)^2}\right)
\nonumber\\
&&+\frac{\left(1+n_{A}\right)\left(2+n_{A}\right)n_{B}}{\lambda_{A}^{+(2)}+\lambda_{B}^{-}}
\left(\frac{1}{\lambda_{A}^{+}}+\frac{1}{\lambda_{A}^{+}+\lambda_{B}^{-}}\right)^2
\nonumber \\
&&+\frac{\left(n_{A}-1\right)\left(1+n_{B}\right)n_{A}}{\lambda_{A}^{-(2)}+\lambda_{B}^{+}}
\left(\frac{1}{\lambda_{A}^{-}}+\frac{1}{\lambda_{A}^-+\lambda_{B}^{+}}\right)^2
\nonumber \\
&&\left.+\frac{\left(1+n_{A}\right)^{2}\left(1+n_{B}\right)}{\left(\lambda_{A}^{+}\right)^{2}\lambda_{B}^{+}}\right] \,
\label{2}
\end{eqnarray}
and analogous for $\alpha_{2BB}^{(2)}$ with the indices $A$ and $B$ interchanged.
\begin{figure}[t]
\includegraphics[width=0.45\textwidth]{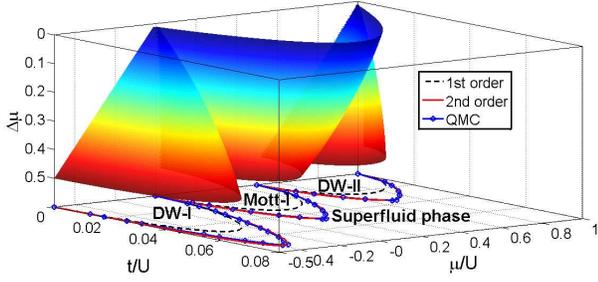}
\caption{(Color online) The quantum phase diagram of a bosonic
2$d$ square superlattice in the whole parameter region
from first hopping order GEPLT, and the phase diagram projected in
the $t$-$\mu$ plane comparing the quantum Monte Carlo simulation
(dotted line), the 1st order (dashed line) and the 2nd order
(solid line) analytic results at $\Delta\mu/U=0.5$.}
\label{figure2}
\end{figure}
\subsection{Quantum Phase Diagram}
In order to got the whole quantum phase diagram, we study at first the
$t=0$ contribution of the effective
potential in Eq.~(\ref{ga}), i.e. $F_0(t=0)=N_{s} E^{(0)}\left(n_{A},n_{B}\right)$
with Eq.~(\ref{E}). We assume $\Delta \mu$ to be in the region
of $[0,U)$. Similar to the normal Bose-Hubbard model, there exist
Mott insulator phases (Mott-$n$), which are characterized by the uniform filling
$n_{A}=n_{B}=n$. However, due to the local offset $\Delta \mu$,
this happens only in the regions
\begin{eqnarray}
\mbox{Mott-$n$:} \quad U\left(n - 1 \right) < \mu < U n -\Delta \mu \, .
\label{C1}
\end{eqnarray}
On the other hand the density wave phases (DW-$n$) break
the translational order as they have the property
$n_{B}=n_A-1$, $n_{A}=n$, yielding the filling factor
$n+1/2$, and minimize the free energy in the other regions
\begin{eqnarray}
\mbox{DW-$n$:} \quad (n-1)U-\Delta \mu<\mu<(n-1)U \, .
\label{C2}
\end{eqnarray}
Hence, depending on the chemical potential offset $\Delta \mu$, we find a natural competition
between Mott phases and density wave phases.
\begin{figure}[t]
\includegraphics[width=0.5\textwidth]{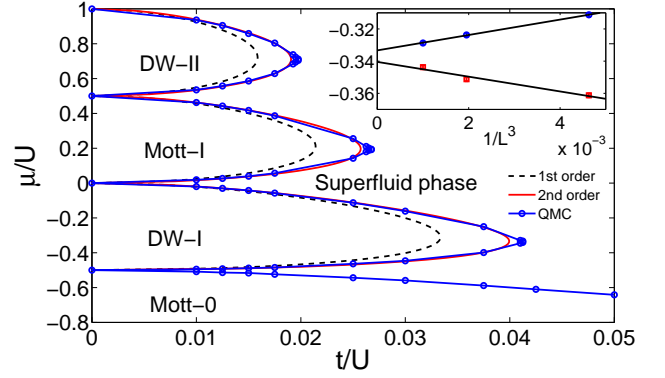}
\caption{(Color online) The quantum phase boundaries of a bosonic
3$d$ cubic superlattice at $\Delta\mu/U=0.5$ which is obtained by
1st order (dashed line) and 2nd order (solid line) generalized
effective Landau potential theory, and the quantum Monte Carlo
simulation (dotted line) in the thermodynamic limit. Inset:
Finite-size scaling of the critical  points of DW-I at $t/U=0.04125$.}
\label{figure3}
\end{figure}

Turning on the hopping processes, the quantum fluctuations will
melt the different insulating phases, and the critical lines are
determined in second hopping order by Eq.~(\ref{tc}) after
substituting the respective strong-coupling coefficients
$\alpha^{(n)}_{2ll'}$ from Eqs.~(\ref{0})--(\ref{2}). The
resulting quantum phase diagram for square and cubic
superlattices are shown in Fig.~\ref{figure2} and
Fig.~\ref{figure3}, respectively.

\begin{figure}[t]
\includegraphics[width=0.5\textwidth]{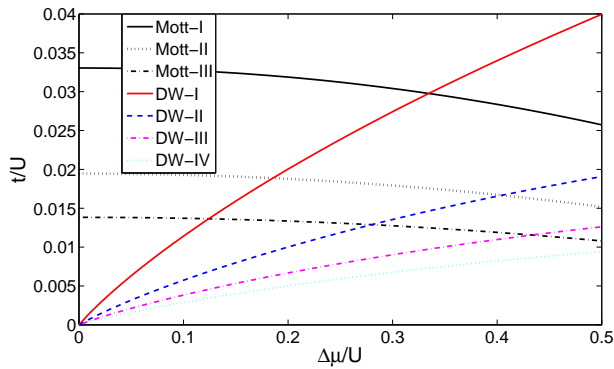}
\caption{(Color online) Maxima of Mott-$n$ and DW-$n$ lobes as a function of $\Delta \mu$ for a 3$d$ cubic
superlattice.}
\label{figure4}
\end{figure}

From the GEPLT calculation we find for the special case $\Delta
\mu=0$ that the Mott-1 lobe coincides with the single component method from
Ref.~[\onlinecite{pelster2}] as expected.
In addition, when $\Delta\mu$ is larger than zero,
the DW phase appears, and its region increases with increasing $\Delta\mu$, whereas the region of the Mott phase decreases
correspondingly. This is a direct consequence of the translational
symmetry breaking due to the superlattice structure. Furthermore, this observation is confirmed by
a direct comparison of the lobe maxima according to Fig.~\ref{figure4}, where
the tips of the Mott lobes decrease with increasing $\Delta\mu$, whereas
the tips of the DW lobes increase.

Comparing the quantum phase diagram in different dimensions, we
notice that not only the lobes of the Mott insulators but also the
DW phases are smaller in three than in two dimensions, which
indicates that the dimensionality has a similar effect on both
incompressible phases. In addition, in order to check the accuracy
of GEPLT, we have developed a quantum Monte Carlo algorithm on the basis of a
stochastic series expansion \cite{zhang,sse1,sse2,wen1,wen2,wen3,xuefeng}
and performed high-precision simulations
for different superlattice systems. After finite-size
scaling up to 144 sites in 2d and 1000 sites in 3d shown in the
inset of Fig.~\ref{figure3}, we obtained the corresponding quantum phase diagrams in the
thermodynamic limit. Their good match with GEPLT indicates the
efficiency of our algorithm.

In principle, it would also be quite interesting to investigate in detail the question which critical exponents occur for the
lobes of the Mott insulators and DW phases.
To this end we refer first of all to the usual Bose-Hubbard model where,
concerning the static critical exponents, it does not matter
at which point the lobe is crossed, while the dynamic critical exponent depends on whether the crossing occurs at the tip of the lobe or whether it is crossed
somewhere else \cite{Fisher:PRB89}. Furthermore, critical exponents are trivial in 3d as they coincide with mean-field values, whereas they are nontrivial in 2d as they
deviate from mean-field theory \cite{Fisher:PRB89}. It would be quite challenging to transfer the techniques of Refs. \cite{exponents1,exponents2}
for determining critical exponents from the normal lattice systems to superlattices,
but we consider this topic to be more suitable for a future research work.

Note that the
Bose-Hubbard model in the superlattice system can also be analyzed
by using the decoupled mean-field
theory \cite{vezzani,hou,shuchen,das1}, where the quantum phase
boundary turns out to agree with our first-order hopping result.
Therefore, we compare in the next section the advantages and
disadvantages of GEPLT with this decoupled mean-field theory.

\section{Comparison with decoupled Mean-Field Theory}\label{AD}
In order to treat a superlattice system with the decoupled
mean-field theory, the operators $\hat{a}^{\dag}_i$($\hat{a}_i$)
are decomposed into the mean fields $\psi_i^{*}$ ($\psi_i$), which
are identified with the order parameters, and the remaining
operators $\delta\hat{a}^{\dag}_i$($\delta\hat{a}_i$), which
describe the quantum fluctuations around the mean fields. Then,
after neglecting second order terms of the quantum fluctuations
and assuming due to homogeneity that the order parameters are equal in the same subsystem,
the Bose-Hubbard Hamiltonian (\ref{bh}) decouples into a
mean-field Hamiltonian on two subsystems \cite{vezzani,hou,shuchen,das1}:
\begin{eqnarray}
\hat{H}_{\rm MF}&=&\hat{H}_{0} +tz \Big( \psi_{A}^{*} \psi_{B}+
\psi_{B}^ {*} \psi_{A} \nonumber \\
&&
- \hat{a}_{A}^{\dagger} \psi_{B}-\hat{a}^{\dagger}_{B} \psi_{A}
-\hat{a}_{A} \psi_{B}^{*} -\hat{a}_{B} \psi_{A}^{*}
 \Big)
\, ,
\end{eqnarray}
with $\hat{H}_{0}$ from (\ref{HH0}).
Because the order parameters are tiny near the boundary of the
second-order phase transition, the corresponding free energy can be Taylor
expanded with respect to both order parameters
\begin{eqnarray}
F_{\rm
MF}&=&f_{0}+f_{2A}\left|\psi_{A}\right|^{2}+f_{2B}\left|\psi_{B}\right|^{2}
+f_{2AB}\psi_{A}\psi_{B}^{*}\nonumber\\
&&+f_{2BA}\psi_{B}\psi_{A}^{*}+ ... \, ,
\end{eqnarray}
where the leading term $f_0$ is equal to the leading term
$F_0(t)$ of GEPLT at $t=0$. Thus, from the previous analysis on $F_0(t=0)=N_{s} E^{(0)}\left(n_{A},n_{B}\right)$
with Eq.~(\ref{E}), we obtain the restrictions (\ref{C1}) and (\ref{C2}) for the chemical potential
in the Mott insulator and density wave
phases, respectively.
As we only consider the system at zero temperature, the free
energy is equivalent to the ground-state energy, which can be
calculated perturbatively in the occupation number representation.
With this we get the second-order coefficients
\begin{eqnarray}
&f_{2A}&=t^{2}z^{2}\left[\frac{n_{B}}{U\left(n_{B}-1\right)-\mu}+\frac{n_{B}+1}{\mu-Un_{B}}\right]
\, ,
\\
&f_{2B}&=t^{2}z^{2}\left[\frac{n_{A}}{U\left(n_{A}-1\right)-\mu-\Delta
\mu}+\frac{n_{A}+1}{\mu+\Delta \mu-Un_{A}}\right]\,. \nonumber
\end{eqnarray}
With the conditions (\ref{C1}) and (\ref{C2}) we find for both second-order
derivatives the inequalities
\begin{eqnarray}
\left. \frac{\partial^{2} F_{\rm MF}}{\partial \psi_{A}\partial
\psi_{A}^{*}}\right|_{\psi_{A},\psi_{B}=0}&=&f_{2A}<0 \, ,
\nonumber \\
\left. \frac{\partial^{2} F_{\rm MF}}{\partial \psi_{B}\partial
\psi_{B}^{*}}\right|_{\psi_{A},\psi_{B}=0}&=&f_{2B}<0 \, .
\end{eqnarray}
This contradicts with the minimum condition which requires that both
second-order derivatives  are
positive at $\psi_{A}=\psi_{B}=0$. We consider this to be
a general problem of the multi-component decoupled mean-field theory,
because it also happens in other systems such as Kagome and
triangular systems. Note that it can be shown that a single-component
mean-field theory does not have this minimum problem.

In order to proof that the GEPLT does not suffer from such
a problem, we conclude at first from Eq.~(\ref{relation})
\begin{equation}
\begin{aligned}
&\left|
\begin{array}{ccc}
a_{2AA} & a_{2AB} \\
a_{2BA} & a_{2BB}
\end{array}
\right| = \frac{1}{\left|
\begin{array}{ccc}
-c_{2AA} & -c_{2AB} \\
-c_{2BA} & -c_{2BB}
\end{array}
\right|}  \\
&=\frac{1}{(\sqrt{c_{2AA}c_{2BB}}+c_{2AB})(\sqrt{c_{2AA}c_{2BB}}-c_{2AB})}\,.
\end{aligned}\label{dom}
\end{equation}
Considering, for instance, the first-order result, we have in the
Mott lobe $c_{2AB}^{(1)}=-tz\alpha_{2AA}^{(0)}\alpha_{2BB}^{(0)}<0$,
$c_{2AA}^{(0)}=\alpha_{2AA}^{(0)}<0$, $c_{2BB}^{(0)}=\alpha_{2BB}^{(0)}<0$, so we
get up to first order $\sqrt{c_{2AA}c_{2BB}}-c_{2AB}>0$. Considering
$\sqrt{c_{2AA}c_{2BB}}+c_{2AB}=0$ is the phase boundary and $t$ is
decreasing from the superfluid phase to the insulator phase, the denominator
of Eq.~(\ref{dom}) is positive in the insulator lobe which means
\begin{eqnarray}
\left|
\begin{array}{ccc}
a_{2AA} & a_{2AB} \\
a_{2BA} & a_{2BB}
\end{array}
\right|>0
\end{eqnarray}
and
\begin{eqnarray}
 a_{2AA}=\frac{-c_{2BB}}{\left|
\begin{array}{ccc}
-c_{2AA} & -c_{2AB} \\
-c_{2BA} & -c_{2BB}
\end{array}
\right|}>0\,.
\end{eqnarray}
For the same reason, $a_{2BB}$ is also positive. Thus, the
effective potential is really a local minimal at the zero point.
Thus, in comparison with the decoupled mean-field approach
GEPLT has the decisive advantage to be consistent for
superlattice systems.

Another advantage of our method is its higher accuracy. In
comparison with quantum Monte Carlo simulations, the error of the
GEPLT is less than $3 \%$ in second hopping order. And, according
to our knowledge, such high accuracy is hard to reach by using
other analytic methods. It can only be surpassed by higher hopping
orders which could be evaluated via the process-chain approach of
Refs.~[\onlinecite{Eckardt,Holthaus5,Holthaus6,Teichmann,Holthaus7}].

However, the GEPLT also has its disadvantages. For the
two-dimensional square superlattice system we can not
get the full lobe of the phase boundary of the DW-I phase in the
parameter range $\Delta \mu<0.35 U $ in second hopping order,
because the radicand of the square root in the phase boundary
Eq.~(\ref{tc}) becomes negative in second hopping order. We
suspect that this is an artifact of truncating the hopping
expansion at second order and expect that this could be corrected
by obtaining higher hopping orders.
\section{Conclusion and Outlook}\label{CO}
In this paper, we extended the single-component effective
potential Landau theory to the general case of a  multi-component
GEPLT method. In order to include several order parameters, we
introduced the source vectors into the general multi-component
Bose-Hubbard Hamiltonian. After performing the Legendre
transformation of the free energy, we obtained a generalized
effective potential, which can be determined in an expansion in
hopping matrix elements.  This method can be applied to the
bosonic square and cubic superlattice systems yielding
high accuracy results for the phase diagrams in second order
hopping compared to QMC simulations. Apart from the Mott insulator
phases, we also found competing DW phases with fractional filling
factors which are induced by the translational symmetry breaking
of the superlattice system. The dimensionality has a similar
effect on the Mott insulator and the DW phases. Compared  with the
decoupled mean-field theory, the GEPLT has a higher accuracy and
does not suffer from the local minimum problem. However, GEPLT
also has a problem in calculating the whole quantum phase diagram
for the DW-I phase which should be solved by considering higher
order hopping corrections.

As the GEPLT turned out to be a general method for detecting second-order
quantum phase transitions in a system with multi order parameters, we
think it is also suitable for frustrated superlattice systems,
such as the triangular and the Kagome lattice. Since the
supersolid-solid phase transition for hard-core bosons is
found to be of second order in the triangular
lattice \cite{xuefeng}, the GEPLT introduces a promising way to detect the quantum phase
transition in both positive and negative hopping process regions.
Furthermore, extending this work for
finite temperatures and investigating the universal properties near the
quantum phase boundary, are certainly worth
for more detailed studies in the future.
\begin{acknowledgments}
X.F. Zhang acknowledges inspiring discussions with Y.C. Wen on
numerical simulations and the physical understanding of the
superlattice system. T. Wang thanks for the financial support from the
Chinese Scholarship Council (CSC).  This work is also supported by the
German Research Foundation (DFG) via the Collaborative Research Center
SFB/TR49.
\end{acknowledgments}
\begin{appendix}
\section{Strong-Coupling Peturbation Theory} \label{AA}
The perturbative coefficients
$\alpha_{2ll'}^{(n)}$ follow at zero temperature
from applying Rayleigh-Schr\"{o}dinger perturbation theory using a suitable
diagrammatic representation \cite{pelster2}.
By denoting the creation
(annihilation) operator with an arrow line pointing into (out of)
the site, each perturbative contribution of $\alpha_{2ll'}^{(n)}$
can be sketched as an arrow-line diagram which is composed of $n$
oriented internal lines connecting the vertices and two external
arrow lines. The vertices in the diagram correspond to the
respective lattice sites, oriented internal lines stand for the
hopping process between sites, and the two external arrow lines
are representing creation and annihilation operators,
respectively. Table~\ref{table-1} presents all non-vanishing
arrow-line diagrams as well as the associated multiplicities
$\alpha_{2ll'}^{(n)}$ up to the second hopping order.

Note that the arrow-line diagrams only depict the possible hopping
processes. In order to determine each non-zero perturbative term
$\alpha^{(n)}_{2ll'}$, we also invoke a line-dot diagrammatic
representation which has been worked out for a single component
method in Ref.~[\onlinecite{pelster2}]. To this end we consider the general situation
that a Hamiltonian $\hat{H}$ decomposes into
an unperturbed term $\hat{H}_0$ and
two perturbation terms $\hat{V}$, $\hat{W}$, i.e.
\begin{eqnarray}
\hat{H}=\hat{H}_{0}+\lambda \hat{V}+\delta \hat{W}\,,
\end{eqnarray}
where $\lambda$ and $\delta$ are small parameters. We then
calculate the zero-temperature free energy by using perturbation theory. Each
term is related to several line-dot diagrams which stem from the
following rules:
\begin{itemize}
\item The dots labeled by $1$ and
$2$ represent the perturbative terms $\hat{V}$ and $\hat{W}$, respectively.
\item The internal lines connecting two adjacent dots are associated
with the factor
\begin{eqnarray*}
\sum_{m\neq n}\frac{1}{\left( E_{n}^{\left(
0\right) }-E_{m}^{\left( 0\right) }\right) ^{p}}| \Psi_{m}^{\left(
0\right) }\rangle\langle \Psi_{m}^{\left( 0\right) }|\,,
\end{eqnarray*}
where the
ground state of $\hat{H}_0$ is $|\Psi_{n}^{\left(
0\right)}\rangle$ with the energy $E_n^{(0)}=\langle \Psi_{n}^{\left(
0\right) }|\hat{H}_0|\Psi_{n}^{\left( 0\right)}\rangle$ and
$|\Psi_{m}^{\left( 0\right)}\rangle$ represents an excited state with
the energy $E_m^{(0)}$, whereas $p$ denotes the number of lines connecting
two given consecutive dots.
\item $\langle \Psi_{n}^{\left(  0\right) }\vert $ and $\vert
\Psi_{n}^{\left( 0\right) }\rangle $ are denoted by
left-external and right-external lines, respectively, so in the
diagrammatic representation of $E_{n}^{\left(  i\right)  }$ there
are some graphs which consist of $s$ disconnected parts.
The weight of these graphs has to be multiplied by the sign $\left(  -1\right) ^{s-1}$.
\end{itemize}
\begin{table}[t]
\center
\begin{tabular}
[c]{|c|c|c|c|} \hline & & & \\
$\alpha_{2AA}^{(0)}$ & $\includegraphics[width=1.5cm]{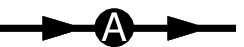}$ &
$\alpha_{2BB}^{(0)}$ & $\includegraphics[width=1.5cm]{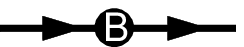}$
\\
& & &  \\
\hline
& & &\\
$\alpha_{2AB}^{(1)}$ & \hspace{0.25cm}
$z\includegraphics[width=2.2cm]{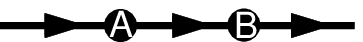}$ \hspace{.2cm}&
$\alpha_{2BA}^{(1)}$  &$z\includegraphics[width=2.2cm]{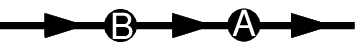}$\\
& & &\\
\hline
& \multicolumn{3}{|l|}{} \\
$\alpha_{2AA}^{(2)}$&\multicolumn{3}{|c|}{$z\includegraphics[width=2.5cm,height=1.5cm]{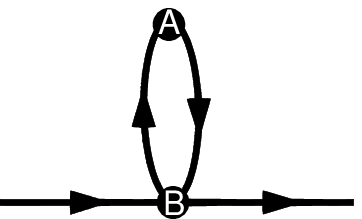}+z(z-1)\includegraphics[width=3.3cm]{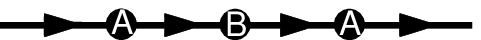}$}\\
& \multicolumn{3}{|l|}{} \\
\hline
& \multicolumn{3}{|l|}{} \\
$\alpha_{2BB}^{(2)}$&\multicolumn{3}{|c|}{$z\includegraphics[width=2.5cm,height=1.5cm]{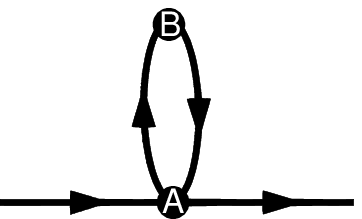}+z(z-1)\includegraphics[width=3.3cm]{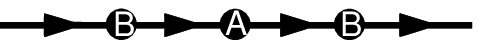}$}\\
& \multicolumn{3}{|l|}{} \\
\hline
\end{tabular}
\caption{Diagrammatic arrow line expressions of the non-vanishing
elements $\alpha_{2ll'}^{(n)}$ including their multiplicities for
a square and cubic superlattice up to second hopping
order. The coordinate number $z$ is $2d$ for a $d$ dimensional
hypercubic lattice.} \label{table-1}
\end{table}
With these rules, we obtain within the line-dot representation
the perturbative expansion
\begin{eqnarray}
&& E_{n}= E_{n}^{(0)}+\lambda
\raisebox{-0.05cm}{\hspace{0.05cm}\includegraphics[width=1cm]{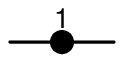}\hspace{0.05cm}}+\delta\raisebox{-0.05cm}{
\hspace{0.05cm}\includegraphics[width=1cm]{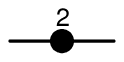}\hspace{0.05cm}}+\lambda
\delta \left(\raisebox{-0.05cm}{\hspace{0.05cm}\includegraphics[width=1.4cm]{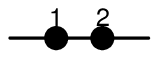}\hspace{0.05cm}} \right.\nonumber \\
&&\left.+\raisebox{-0.05cm}{\hspace{0.05cm}\includegraphics[width=1.4cm]{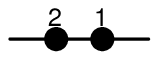}\hspace{0.05cm}}\right)
+\lambda^{2}\raisebox{-0.05cm}{\hspace{0.05cm}\includegraphics[width=1.4cm]{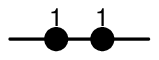}\hspace{0.05cm}}
+\delta^{2}\raisebox{-0.05cm}{\hspace{0.05cm}\includegraphics[width=1.4cm]{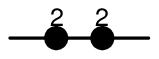}\hspace{0.05cm}}
\\
&&+\lambda^{2}\delta
\left(\raisebox{-0.05cm}{\includegraphics[width=1.8cm]{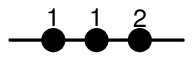}}+
\raisebox{-0.05cm}{\includegraphics[width=1.8cm]{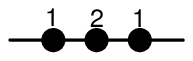}}+\raisebox{-0.05cm}{\includegraphics[width=1.8cm]{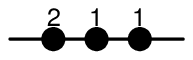}}
\right.
\nonumber\\
&&\left.+\raisebox{-0.5cm}{\includegraphics[width=1.8cm]{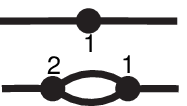}}+\raisebox{-0.5cm}{\includegraphics[width=1.8cm]{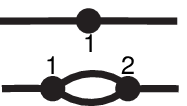}}
+\raisebox{-0.5cm}{\includegraphics[width=1.8cm]{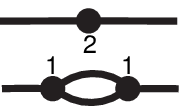}}\right)+\cdots\,.
\nonumber
\end{eqnarray}
In our concrete case of the square and cubic
superlattice with the Hamiltonian in Eq.~(\ref{bh}), the
unperturbed Hamiltonian is given by
\begin{eqnarray}
\hat{H}_{0}&=&\frac{U}{2} \hat{n}_{A} \left(\hat{n}_{A}-1
\right)+\frac{U}{2} \hat{n}_{B} \left(\hat{n}_{B}-1 \right)\nonumber \\
&&-(\mu+\Delta \mu)
\hat{n}_{A} -\mu \hat{n}_{B} \,,
\label{HH0}
\end{eqnarray}
yielding the unperturbed energies (\ref{E}), whereas both the hopping and current terms are treated as a perturbation.
Thus this leads to
the arrow diagrams within the coefficient $\alpha_{2ll'}^{(n)}$,
which  can now be represented in terms of respective line-dot diagrams.
Note that each term $\alpha _{2ll'}^{(n)}$ consists of exactly one creation
operator (associated with $J_{i}$), one annihilation operator
(associated with $J_{i}^{*}$), and $n$ hopping operators
(associated with $t^n$).
For each arrow-line diagram we have to draw all possible
topologically different line-dot diagrams. The sum of all these
line-dot diagrams then gives the corresponding result. For example, the equation
\begin{eqnarray}
\label{example}
\raisebox{-0.1cm}{\includegraphics[width=2.0cm]{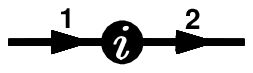}}=\raisebox{-0.05cm}{\includegraphics[width=1.5cm]{linedot3.eps}}+
\raisebox{-0.05cm}{\includegraphics[width=1.5cm]{linedot4.eps}} \, ,
\end{eqnarray}
where $i$ inside the dot stands for a particular sublattice, expresses
exemplary how to transfer an arrow-line diagram into its line-dot
representation. Following these steps one obtains for the respective
coefficients $\alpha_{2ll'}^{(n)}$
the results (\ref{0})--(\ref{2}), where the abbreviations (\ref{E}), (\ref{P}) are used.
\end{appendix}

\end{document}